\documentclass[aps,prb,preprint,superscriptaddress,amsmath,amssymb]
{revtex4-1}
\usepackage{graphicx}
\usepackage{dcolumn}
\usepackage{bm}

\newcommand{\ks} {{\bf k}}

\newcommand{\pc} {{\bf P}}
\newcommand{\rs} {{\bf r}}

\newcommand{\es} {{\bf e}}

\newcommand{\ec} {{\bf E}}
\newcommand{\gc} {{\bf G}}
\newcommand{\jc} {{\bf J}}

\begin{document}


\title{Femtosecond optical breakdown in silicon}

\author{Tzveta Apostolova}

\address{Institute for Nuclear Research and Nuclear Energy, Bulgarian Academy of Sciences, Tsarigradsko chausse 72, 1784 Sofia, Bulgaria}
\address{Institute for Advanced Physical Studies,
New Bulgarian University, 1618 Sofia, Bulgaria}

\author{Boyan Obreshkov}
\address{Institute for Nuclear Research and Nuclear Energy, Bulgarian Academy of Sciences, Tsarigradsko chausse 72, 1784 Sofia, Bulgaria}

\begin{abstract}

We investigate photoinization, energy deposition, plasma formation and the ultrafast optical breakdown in crystalline silicon irradiated by intense near-infrared laser pulses with pulse duration $\tau \le $ 100 fs.
The occurrence of high-intensity breakdown was established by the sudden increase of the absorbed laser energy inside the bulk, which corresponds to threshold energy fluence $\Phi_{th} > $ 1 J/cm$^2$. The optical breakdown is accompanied by severe spectral broadening of the transmitted pulse. For the studied irradiation conditions, we find that the threshold fluence increases linearly with the increase of the pulse duration, while the corresponding laser intensity threshold decreases. The effect of the high plasma density on the stability of diamond lattice is also examined. For near threshold fluences, when about 5 \% of valence electrons are promoted into the conduction band, the Si-Si bonds are softened and large Fermi degeneracy pressure arises (with pressure up to 100 kbar). The mechanical instability of the diamond lattice suggests that the large number of electron-hole pairs leads directly to ultrafast melting of the crystal structure.

\end{abstract}

\maketitle

\section{Introduction}
\label{S:1}

Ultrafast breakdown resulting in damage of dielectrics has been subject of extensive experimental and theoretical investigations \cite{LIB1997,Du1994, Stuart1996,Kautek1996,Varel1996,Lenzner1998,Pronko1998,Li1999,
vanDriel1987,Apostolova2000,Schaffer2001,Apostolova2002,Gamaly2002,Kudryashov2007,Rajeev2009,Chimier2011,Uteza2011,Grojo2013,Lebugle2014,Sato2015,Zhokhov2018}.  Femtosecond laser-induced breakdown in dielectrics has been modeled in terms of three basic physical processes, cf. e.g. \cite{Stuart1996,Schaffer2001}
: i) multiphoton and tunnel ionization generating free electrons into the conduction band, ii) avalanche ionization involving free carrier absorption followed by impact ionization and iii) transfer of the electronic excitation energy to the lattice, eventually resulting in optical damage after conclusion of the pulse. The relative importance of these mechanisms have been studied by measuring the dependence of the breakdown threshold on the duration $\tau$ of the applied laser pulse \cite{Stuart1996,Du1994,Lenzner1998,Chimier2011,Uteza2011}. Though experimental results for the measured damage threshold agree for picosecond to nanosecond pulses, a controversy exists on the observed scaling trend of the threshold fluence for femtosecond pulses \cite{Du1994,Stuart1996,Lenzner1998}.
Though measurement of the energy fluence threshold has been extended to sub 10-fs laser pulses \cite{Lenzner1998,Chimier2011}, it is difficult to extract information regarding the time scale for transfer of the electronic excitation energy to the lattice. Femtosecond time-resolved pump-probe measurements of optical breakdown thresholds in dielectrics \cite{Li1999} have demonstrated that the energy transfer to the lattice is an ultrafast process that lasts for 100fs.

Femtosecond time-resolved measurements are very effective in probing non-linear photionization and the temporal behavior of the optical breakdown and damage thresholds \cite{Guo2019}. Non-thermal melting of silicon is thought to be a result of promotion of a large fraction of the valence electrons into the conduction band \cite{Shank1983,Hulin1984,Tom1988,Sokol1995,Rousse2001,Sundaram2002,Harb2008}. The high kinetic pressure of the hot electron distribution softens the transverse acoustic phonon at the Brillouin zone boundary, which results in instability of the diamond lattice against shear stress \cite{Stampfli1990,Stampfli1992,Stampfli1999,Silvestrelli1996}. Theoretical modeling of the ultrafast electron dynamics in dielectrics and insulators has been reviewed in Refs.\cite{Gamaly2011,Balling2013}. More recently, density-functional-theory based modeling of ultrafast electron dynamics in silicon were reported \cite{Sato2014p1,Sato2014p2,Schultze2014,Apostolova2020,
Otobe2020,Yamada2021}. The fundamental physical processes involve coupling of the laser energy into the electron subsystem on an attosecond time scale as a part of the subcycle photoionization dynamics and electron-hole pair excitation.  When the density of photoexcited carriers becomes sufficiently large,  collective plasma oscillations in the electron gas emerge. Plasma formation is also accompanied by highly efficient free-carrier absorption and severe spectral broadening of the transmitted pulse. The last effect was first observed in Ref.\cite{Alfano} and later it was also observed in gases, solids and liquids \cite{Couairon2007,Wang2016,Dubietis2019}.  Supercontinuum generation exhibits threshold behavior and occurs  when the incident laser power exceeds the critical power for self-focusing \cite{Dubietis2017}, the characteristic broad pedestal on the blue side of the spectrum of the transmitted pulse is often associated to occurrence of breakdown \cite{Couairon2007,Gaeta2000,Liu2002,Lu2014,Sudrie2002,Couairon2005}. 

In this paper we report results of theoretical calculations of photoionization and energy deposition by intense femtosecond laser pulses, which allows us to determine the optical breakdown threshold in silicon. More specifically we discuss scaling laws of the absorbed energy and photoelectron density as functions of the laser fluence, and obtain the dependence of the  threshold fluence on the pulse length and laser polarization direction. We consider self-broadening and self-phase modulation effects in the laser breakdown plasma. The photoinduced modification of optical properties are presented in terms of frequency-dependent reflectivity and absorption coefficients. Softening of the Si-Si bonds and the emergent mechanical instability of the diamond lattice are also discussed.

\section{Results and discussion}
Details of the static band structure of silicon, theoretical approach and methodology can be found in Refs. \cite{Lagomarsino2016,Apostolova2020}, where we solve self-consistently coupled time-dependent Schrodinger and Maxwell's equations for covalent dielectrics interacting with linearly polarized intense ultrashort laser pulses of near-infrared wavelength. The time-dependent pulsed electric field of a single pulse transmitted in bulk silicon
\begin{equation}
\ec_{{\rm}}(t)=\ec_{{\rm ext}}(t)+\ec_{{\rm ind}}(t),
\end{equation}
is a superposition of the applied laser electric field $\ec_{{\rm ext}}$ and an induced electric field $\ec_{{\rm ind}}(t)=-4 \pi \pc(t)$ as a result of the polarization of the photoexcited solid. We parameterize the applied electric field by a temporary Gaussian function
\begin{equation}
\ec_{{\rm ext}}(t)=\es e^{-\ln(4) t^2/\tau^2} F \cos \omega_L t
\end{equation}
where $\es$ is a unit vector in the direction of
the laser polarization, $\omega_L$ is the laser frequency corresponding to photon energy $\hbar \omega_L$ = 1.55 eV, $\tau$ is the pulse length and $F$ is the electric field strength related to the peak laser intensity by $I=F^2$.

The pulsed electric field ionizes the valence electrons of Si and builds plasma of electron-hole pairs. Once their number becomes sufficiently large (of order of the Avogadro number), an optical breakdown occurs. We determine the breakdown threshold from the fluence energy dependence of the total electronic excitation energy (per unit cell)
\begin{equation}
E_{abs} = \int^{\infty} dt \jc(t) \cdot \ec(t)
\end{equation}
where $\jc(t)=d \pc / dt$ is the photoinduced current.

\subsection{Optical breakdown threshold - fluence, pulse length and polarization dependence}

\begin{figure}[h]
\centering\includegraphics[width=1\linewidth]{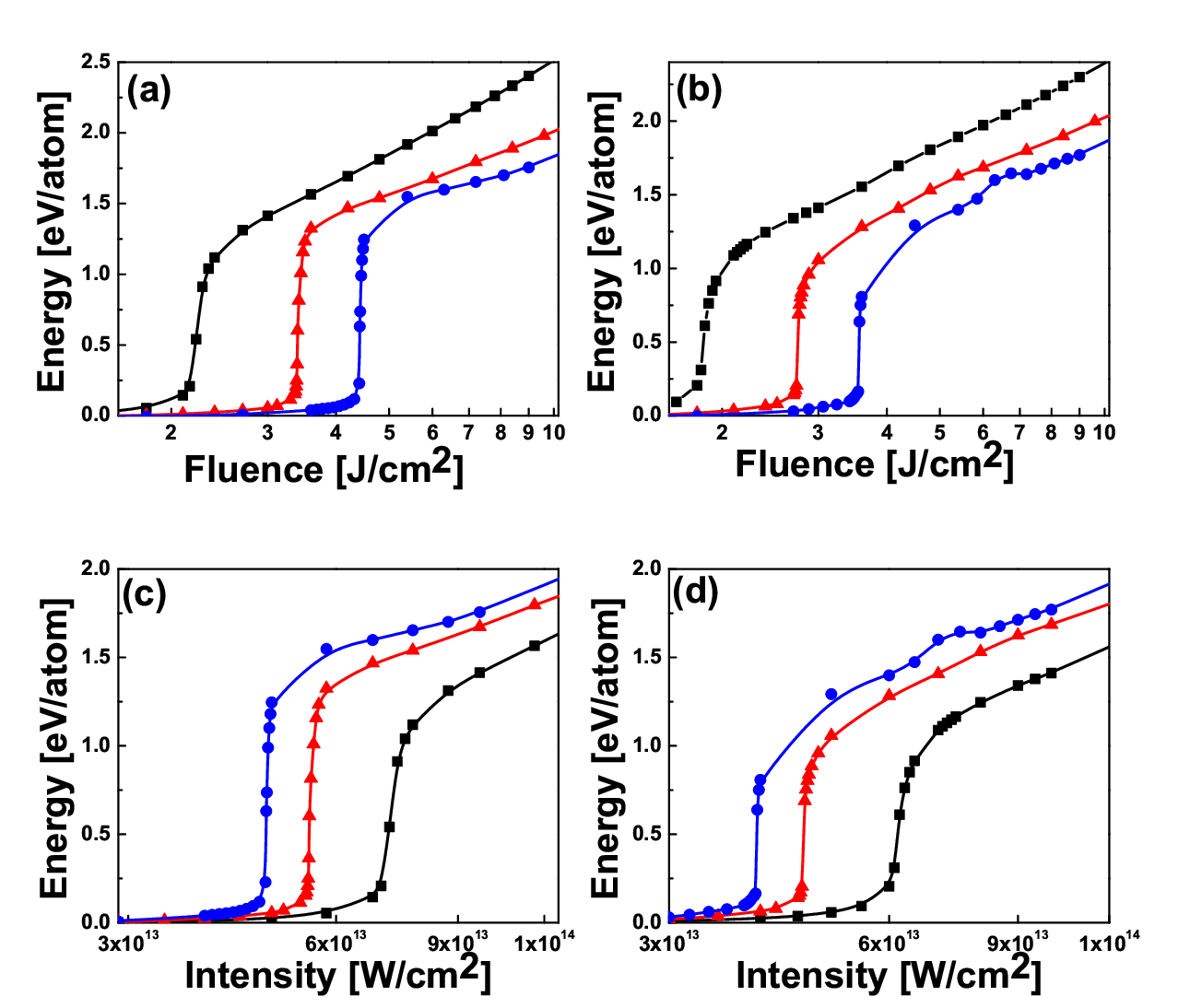}
\caption{Fluence and intensity dependence of the the absorbed energy (in eV/atom) in bulk silicon irradiated by pulsed laser of wavelength 800 nm.  Different symbols in (a-d) designate the laser pulse duration: 30fs (squares), 60 fs  (upper triangles) and 90 fs (circles). The laser is linearly polarized along the [001] crystal direction in Figs. (a),(c), and is polarized   along the [111] direction in Figs.(b) and (d).
}
\label{fig:Fig1}
\end{figure}

 %
 %
As in our earlier investigation \cite{Apostolova2020}, the laser wavelength was set to 800 nm, the peak laser intensity was varied in the range $10^{13}-10^{15}$ W/cm$^2$,  the laser polarization direction was also varied - $\es$ points along the [001] or [111] crystal directions. The crystal was subjected to 30, 60 and 90 fs pulses. Figs.\ref{fig:Fig1}(a-b) show the dependence of the absorbed energy on the laser energy fluence $\Phi=I \tau$. The dependence on the peak laser intensity is also shown in Figs.\ref{fig:Fig1} (c,d).  The low fluence regime with $\Phi \le 1$ J/cm$^2$ is characterized by low conversion efficiency of the laser energy into electronic excitation energy ($E_{abs} \le 0.01$ eV/atom).
For increased fluence, a step-like increase of the absorbed energy by nearly 1 eV/atom occurs (cf. Fig.\ref{fig:Fig1}(a,b)), which we associate with the occurrence of dielectric breakdown in Si.  The fluence threshold depends on the laser polarization direction and on the pulse duration. For laser linearly polarized along the [001]  crystal direction, the threshold fluence $\Phi_{{\rm th}}$ is 1.8,3.1 and 4.3 J/cm$^2$ for 30,60 and 90 fs pulses, respectively. The breakdown threshold decreases for laser polarized along the [111] direction: $\Phi_{{\rm th}}=$ 1.5 (30fs),2.5 (60fs) and 3.4 J/cm$^2$ for the 90fs pulse. For above threshold fluences, energy absorption at near infrared wavelengths continues to increase slowly (following the trend $E_{abs} \sim \Phi^{1/2}$).

\begin{figure}[h]
\centering\includegraphics[width=1\linewidth]{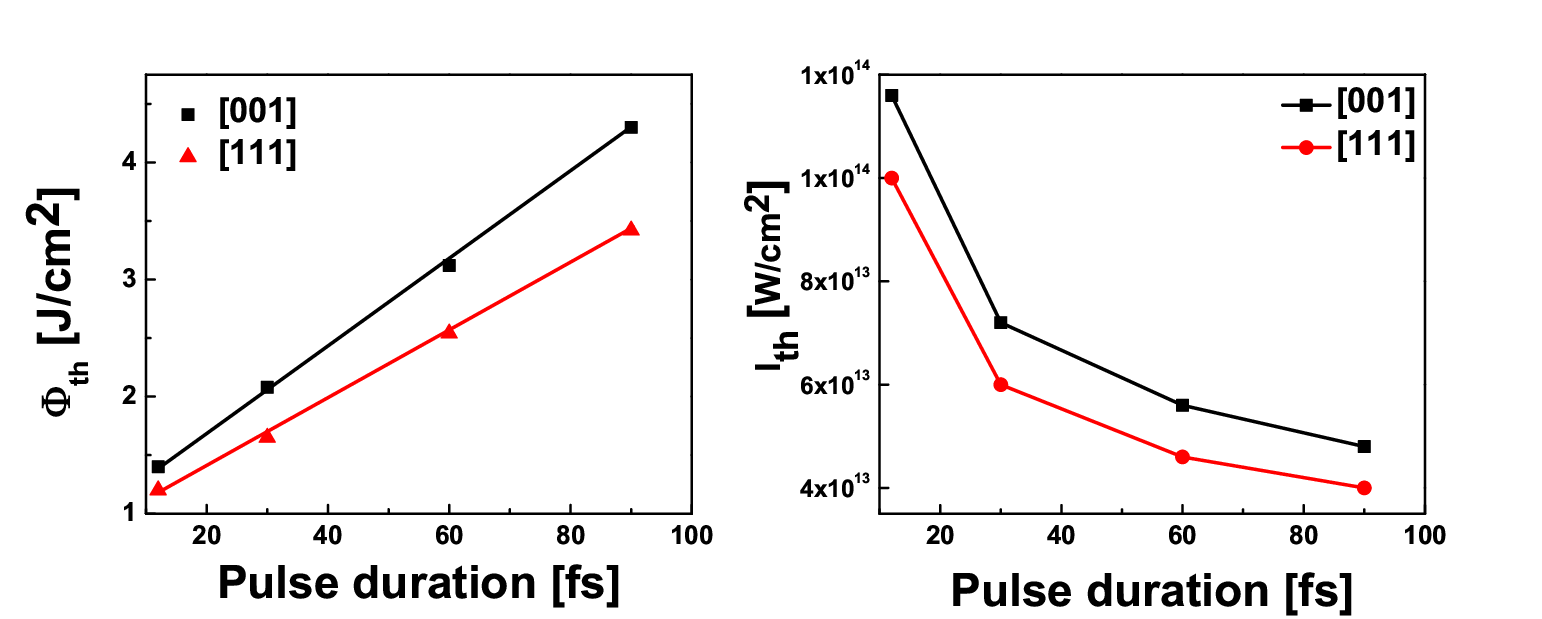}
\caption{Pulse length dependence of the threshold fluence (a) and threshold intensity (b) after irradiation of silicon by a pulsed laser with wavelength 800 nm. Different symbols in (a-b) designate the laser polarization direction: [001] crystal direction (squares) and [111] direction (upper triangles).}
\label{fig:Fig6}
\end{figure}

Independently on the strength of the laser-matter interaction, electrons gain more energy by interacting with longer pulses, such that the energy absorption curves in  Fig.\ref{fig:Fig1} (c-d) do not intersect and remain parallel in the entire laser intensity range. This is a consequence of the uncertainty relation for energy, that is independent on the amount of perturbation. For instance in the  perturbative (low-intensity) regime, the probability for multiphoton transition across a bandgap $\Delta$ is proportional to $\sin^2 [(\Delta-n \hbar \omega_L)\tau/2\hbar]/(\Delta-n \hbar \omega_L)^2$. The distribution is sharply peaked at $\Delta = n \hbar \omega_L$ and has a width is $\sim \hbar/\tau$. As the pulse length increases, the photoionization probability approaches the energy-conserving Dirac delta function $\pi \tau \delta (\Delta - n \hbar \omega_L)$  with a prefactor proportional to $\tau$, hence the absorbed energy also increases linearly with the increase of the pulse length. Fig.\ref{fig:Fig6} (a,b) show the threshold fluence $\Phi_{th}$ and threshold intensity $I_{th}$ as a function of $\tau$, respectively. We find linear dependence  $\Phi_{th}=\Phi_0+I_{\infty}\tau$, where $\Phi_0$ is a threshold fluence extrapolated to zero pulse duration , $I_{\infty}$ is the  threshold intensity for continuous wave laser irradiation ($\tau \rightarrow \infty$). Consequently the threshold intensity decreases with the increase of the pulse length according to $I_{th}=\Phi_0/\tau+I_{\infty}$. These two parameters depend on the laser polarization direction:  $I_{\infty} = 3\times 10^{13}$ W/cm$^2$, $\Phi_0 =0.95$ J/cm$^2$ for [001] polarization direction. $I_{\infty} = 2.8 \times 10^{13}$ W/cm$^2$, $\Phi_0 =0.82$ J/cm$^2$ for [111] polarization direction.
The result agrees only qualitatively with the experimental data of measured damage thresholds in fused silica reported in Refs.\cite{Lenzner1998,Chimier2011}, which exhibit an increasing trend with the increase of the pulse length. However, the linear scaling trend predicted in this work has more profound physical significance, as it shows that the energy uncertainty relation governs the pulse length dependence of the femtosecond optical breakdown threshold.

\begin{figure}[h]
\centering\includegraphics[width=1\linewidth]{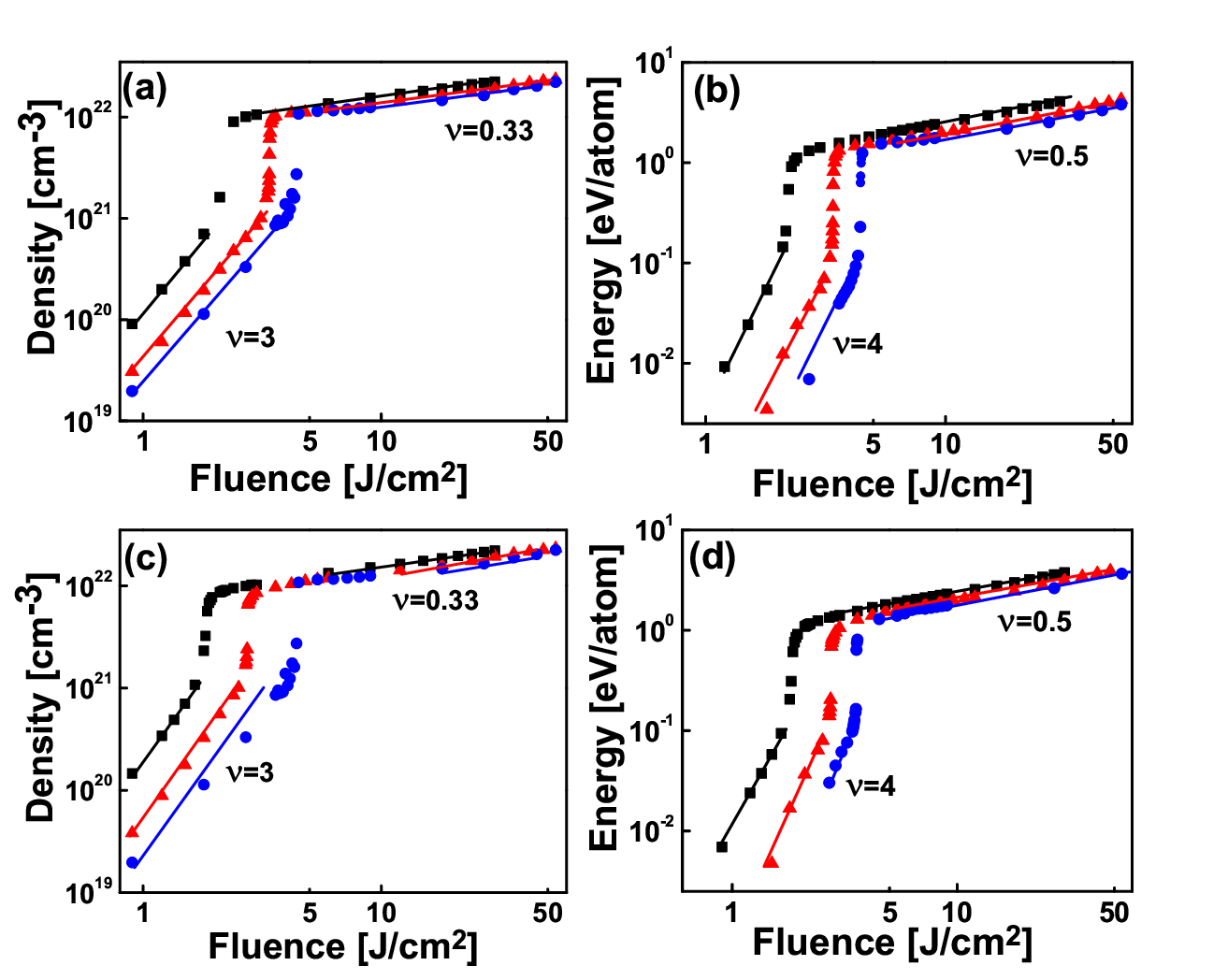}
\caption{Fluence dependence of the the conduction electron density (left panel) and absorbed energy per atom (right panel) in bulk silicon subjected to
pulsed laser irradiation with near infrared wavelength 800 nm. Different symbols in Fig.(a-d) designate the laser pulse duration: 30fs (black squares), 60 fs (upper triangles) and 90 fs (circles). The laser is linearly polarized along the [001] direction in Fig.(a,b) and is polarized along the [111] direction in Fig. (c,d).}
\label{fig:Fig2}
\end{figure}

To derive empirical scaling laws associated with photoionization, in a log-log plot Figs.\ref{fig:Fig2} (a-d) display the photo-electron density and the absorbed laser energy as functions of the laser fluence. The perturbative low-fluence regime is characterized by super-linear scaling trend of the photoelectron density $n_c \sim \Phi^3$ and absorbed energy $E_{abs} \sim \Phi^4$. The scaling laws for energy and density are different as a result of competition between three- and four-photon transitions that are very sensitive to the specific band structure of silicon (cf. below).
For higher laser fluences $\Phi > 1$ J/cm$^2$, a qualitative change of electron dynamics occurs, when both the conduction electron density and absorbed energy undergo a step-like increase at threshold. The discontinous change in conduction electron density is a signature of a first-order phase transition (e.g. melting). Free charge carriers promoted into the conduction band behave similarly to those in a simple metal described by a Drude model. Because their band-averaged effective mass $m^{\ast} \approx 10 m_e$ is large compared to the electron mass, their screened plasma frequency  $\omega_p^2=4 \pi (32/a_0^3) e^2/(m^{\ast} \epsilon_b$) is in the infrared range and nearly matches the laser oscillation frequency (here $\epsilon_b$ is the background dielectric constant of the photoexcited Si).  At this point, resonant energy transfer occurs from the laser pulse to the electrons and the  optical breakdown occurs. The deposited energy at threshold $E_{break} \approx$ 0.2 eV/atom is independent on the parameters of the laser pulse and  slightly exceeds the thermal melting threshold  of silicon (corresponding to temperature $\approx$ 1700 Kelvin).
For above threshold fluences, the electronic excitation energy $E_{abs} > $ 1 eV/atom exceeds a few times the thermal melting threshold and the sub-linear scaling laws $n_c \sim \Phi^{1/3}$  and $E_{abs} \sim \Phi^{1/2}$ are exhibited. The very slow increase of conduction electron density with the increase of fluence is a direct consequence of the exclusion principle:
when large fraction of electrons have already been photoexcited and occupy states in the conduction band,  photoionization into these states becomes unlikely due to Pauli blocking. Hence the density of free charge carriers tends to saturate in the high fluence regime. However the electronic excitation energy does not reach a quasi-saturation limit, instead the photoexcited solid absorbs very efficiently the laser energy with the increased laser fluence. That is because of a  highly-efficient free-carrier absorption: the freed conduction electrons(holes) continue to absorb laser energy via resonant one-photon transitions to vacant higher lying conduction(valence) band states, cf. also Ref. \cite{Apostolova2020}.

The temporal development of the electronic excitation energy for near and above threshold fluences is also shown in
Fig.\ref{fig:Fig5bb} (a-c). For laser fluences slightly above threshold, the laser energy irreversibly transferred to electrons increases in small incremental steps over the full length of the driving pulse. The energy deposition rate increases rapidly with the increased fluence: much larger
portions of energy are transferred to electrons over few half-cycles prior to the pulse peak, whence a highly absorbing state
of silicon is created for near infrared laser wavelengths, with deposited energies exceeding the thermal melting threshold by an order of magnitude.

\begin{figure}[h]
\centering\includegraphics[width=1\linewidth]{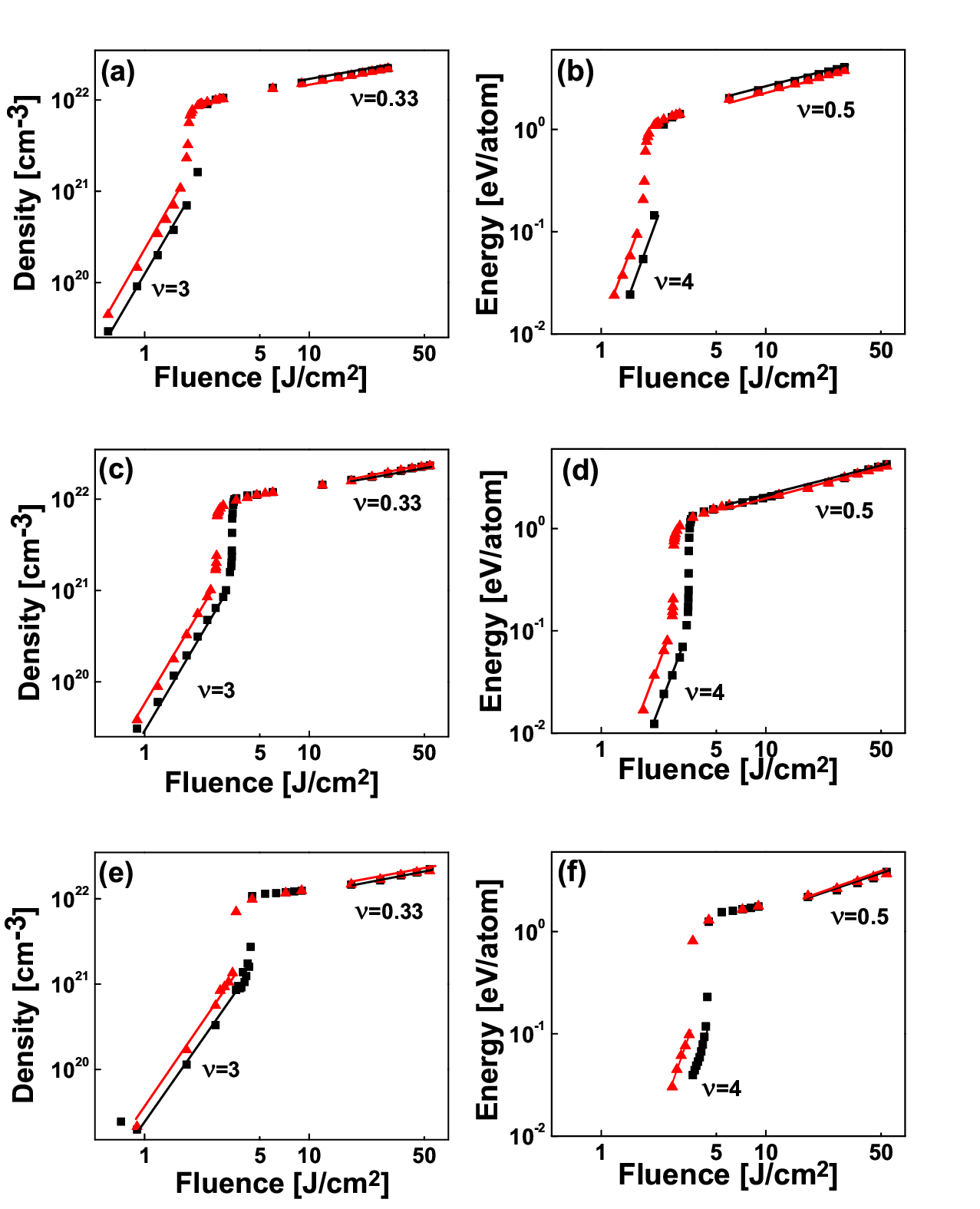}
\caption{Fluence dependence of the conduction electron density (left panel) and the absorbed energy per atom (right panel) in bulk silicon irradiated by pulsed laser with near-infrared wavelength 800 nm. The pulse duration is: 30fs in Figs.(a,b), 60fs in Figs.(c,d) and 90fs in Figs.(e,f).  Different symbols in Fig.(a-f) designate the laser polarization direction: [001] crystal direction (squares) and [111] direction (upper triangles). }
\label{fig:Fig3}
\end{figure}
\begin{figure}[h]
\centering\includegraphics[width=1\linewidth]{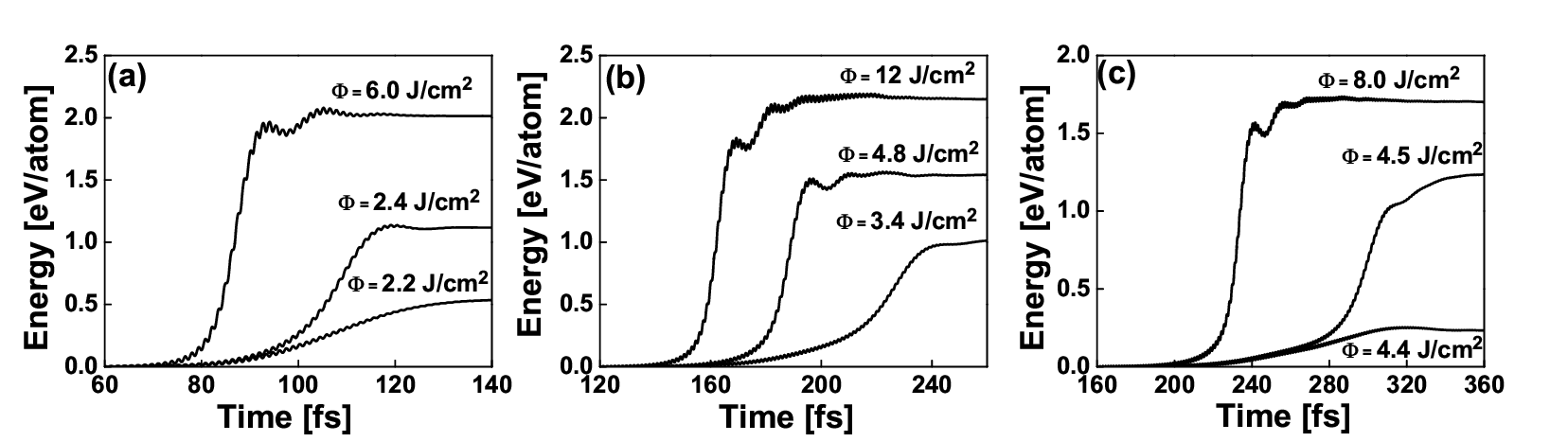}
\caption{Time evolution of the absorbed energy (in eV/atom) in bulk silicon irradiated by pulsed laser with near-infrared wavelength 800 nm. The laser energy fluence (in J/cm$^2$)is indicated above each curve. The pulse duration is  30 fs in Fig.(a),  60 fs in Fig.(b) and 90 fs in Fig.(c).
The laser is linearly polarized along the [001] crystal direction.}
\label{fig:Fig5bb}
\end{figure}

The carrier density and absorbed energy depend also sensitively on the laser polarization direction, cf. Fig.\ref{fig:Fig3} (a-e). The breakdown threshold is lower for a laser pulse linearly polarized along the [111] direction as compared to the [001] direction because in the latter case the laser is aligned with Si-Si bonds, which increases the probability for photoionization.

\subsubsection{Density of conduction electron states}

\begin{figure}[h]
 \centering\includegraphics[width=1\linewidth]{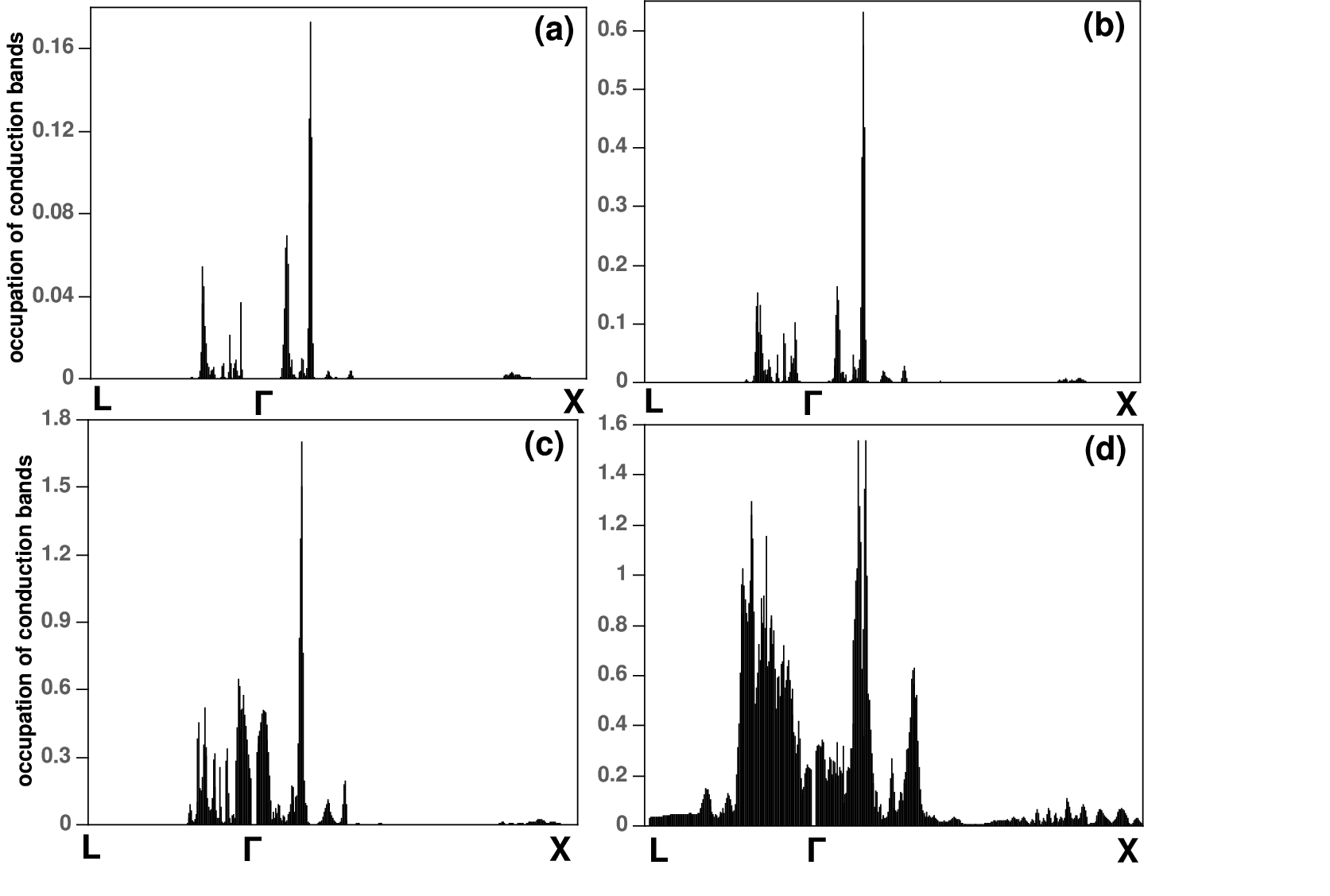}
 \caption{Conduction band fillings in silicon along the $\Delta$ ($\Gamma-X$) and $\Lambda$ ($\Gamma-L$) lines in the Brillouin zone after photo-ionization by 30fs laser pulse with near-infrared wavelength 800 nm, linearly polarized along the [001] crystal direction. The pulse fluence is: 0.9 J/cm$^2$ in Fig.(a), 1.5 J/cm$^2$ in Fig.(b), 2.1 J/cm$^2$ in Fig.(c) and 2.7 J/cm$^2$ in Fig.(d).   }
 \label{fig:focc}
\end{figure}

Specific details on the band structure of Si and its relevance for the excitation mechanism are included in the distribution function of conduction electrons over crystal momenta in the first Brillouin zone after the irradiation by intense 30fs pulse, cf. Figs.\ref{fig:focc}(a-d). In the perturbative low intensity regime with $\Phi=0.9$ J/cm$^2$, shown in Fig. \ref{fig:focc}(a), the distribution consists of narrow lines associated with energy-conserving multiphoton transitions at specific locations in the Brillouin zone. The height of the absorption peaks and thus the probability for photionization increase rapidly with the increased fluence $\Phi=1.5$ J/cm$^2$, specific states along the $\Delta$-line are occupied with high probability. Absorption of laser photons becomes much more frequent for near threshold fluences with $\Phi=2$ J/cm$^2$, shown in Fig.\ref{fig:focc}(c): large fraction of valence electrons are promoted into the conduction band, new structure emerges due to photo-ionization into states near the Brillouin zone center. For above threshold fluences with $\Phi \approx 3$ J/cm$^2$, the distribution broadens in momentum space and more conduction electron states are filled with high probability, cf. Fig. \ref{fig:focc}(d). After that point the capacity of conduction bands is weakened, whence photionization rate decreases and the number of conduction electrons tends to saturate.

\begin{figure}[h]
\centering\includegraphics[width=1\linewidth]{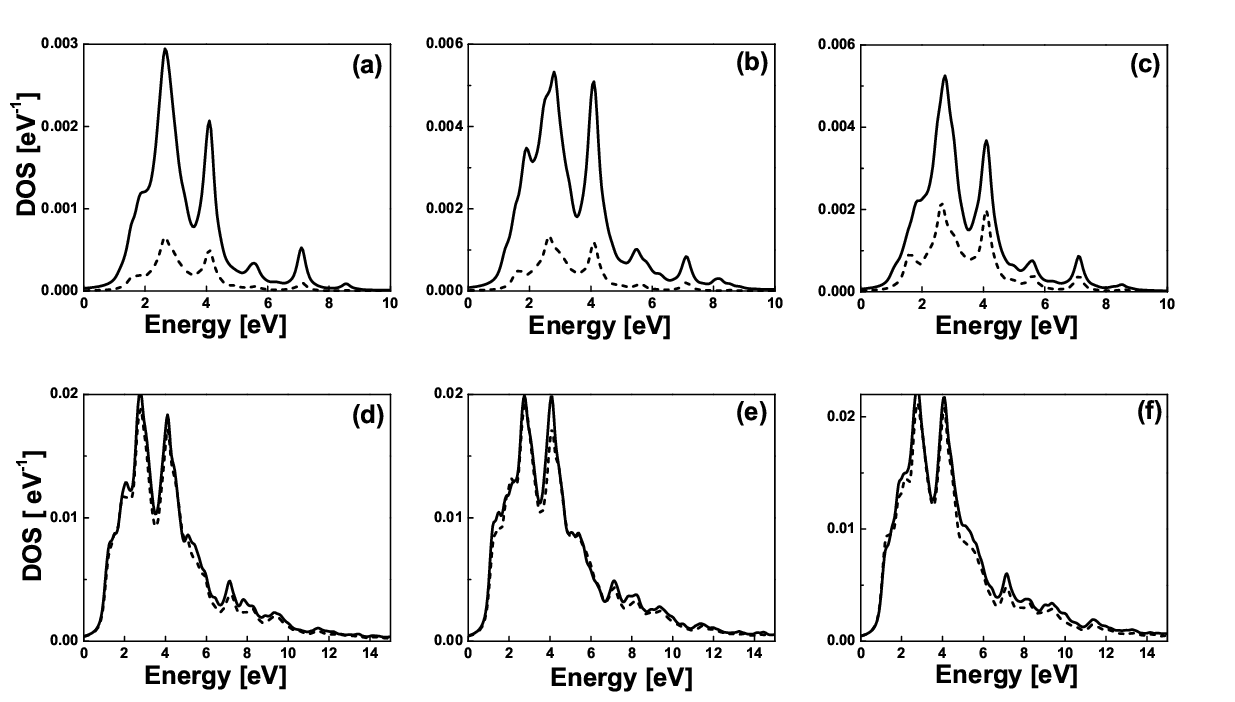}
\caption{Density of conduction electron states per unit cell in bulk silicon after irradiation by a pulsed laser with near-infrared wavelength 800 nm.
The pulse length is 30 fs in Figs.(a,d), 60fs in Figs.(b,e) and 90 fs in Figs.(c,f). The different line types in Fig.(a-f) designate different laser fluences:  In Fig. (a) the laser fluence is 0.9 J/cm$^2$ (dashed line) and 1.5 J/cm$^2$ (solid line); In Fig. (b) - 1.8 J/cm$^2$ (dashed  line) and 2.7 J/cm$^2$ (solid line); In Fig.(c) - 2.7 J/cm$^2$ (dashed line) and 3.4 J/cm$^2$ (solid line). In Fig. (d) - 2.1 J/cm$^2$ (dashed line) and 2.4 J/cm$^2$ (solid line). In Fig. (e) - 3.6 J/cm$^2$ (dashed line) and 4.2 J/cm$^2$ (solid line). In Fig. (f) 5.4 J/cm$^2$ (dashed line) and 6.3 J/cm$^2$ (solid line). The laser is linearly polarized along the [001] crystal direction.}
\label{fig:Fig5}
\end{figure}

The density of conduction electron states after conclusion of the pulse is also shown in Fig.\ref{fig:Fig5}(a-f). In the perturbative regime, cf. Fig.\ref{fig:Fig5}(a-c), the energy distribution is structured into two well-separated broad peaks of nearly equal height with energies 2.5 and 4 eV above the valence band maximum, which are due to energy-conserving 3- and 4-photon transitions associated with energy bands at specific locations in the Brillouin zone. The electron number density and the excitation energy are given by the zeroth and first moments of the electron distribution $n_c(\Phi)=\int d \epsilon \rho(\epsilon,\Phi)$ and $E_{abs}(\Phi)= \int d \epsilon \epsilon \rho(\epsilon,\Phi)$, respectively. Because of the larger spectral weight in the three-photon peak, more electrons are born through the 3-photon energy gaps in Si and thus $n_c \sim \Phi^3$. However more energy is deposited though the 4-photon energy gaps, the higher energy absorption peak contributes more to the absorbed energy and hence $E_{abs} \sim \Phi^4$. The height of the 3-photon and 4-photon absorption peaks increases with the increase of the laser intensity , cf. Fig.\ref{fig:Fig5}(d-f) and a tail of highly energetic photo-electrons emerges. In this high fluence regime, when substantial part of valence electrons are promoted into the conduction band, the role of Pauli's exclusion principle becomes prominent: the density of states exhibits a saturation trend as the laser can not ionize electrons into already occupied states in conduction bands.

\subsection{Optical properties of the laser-excited silicon and supercontinuum generation}

\begin{figure}[h]
\centering\includegraphics[width=0.5\linewidth]{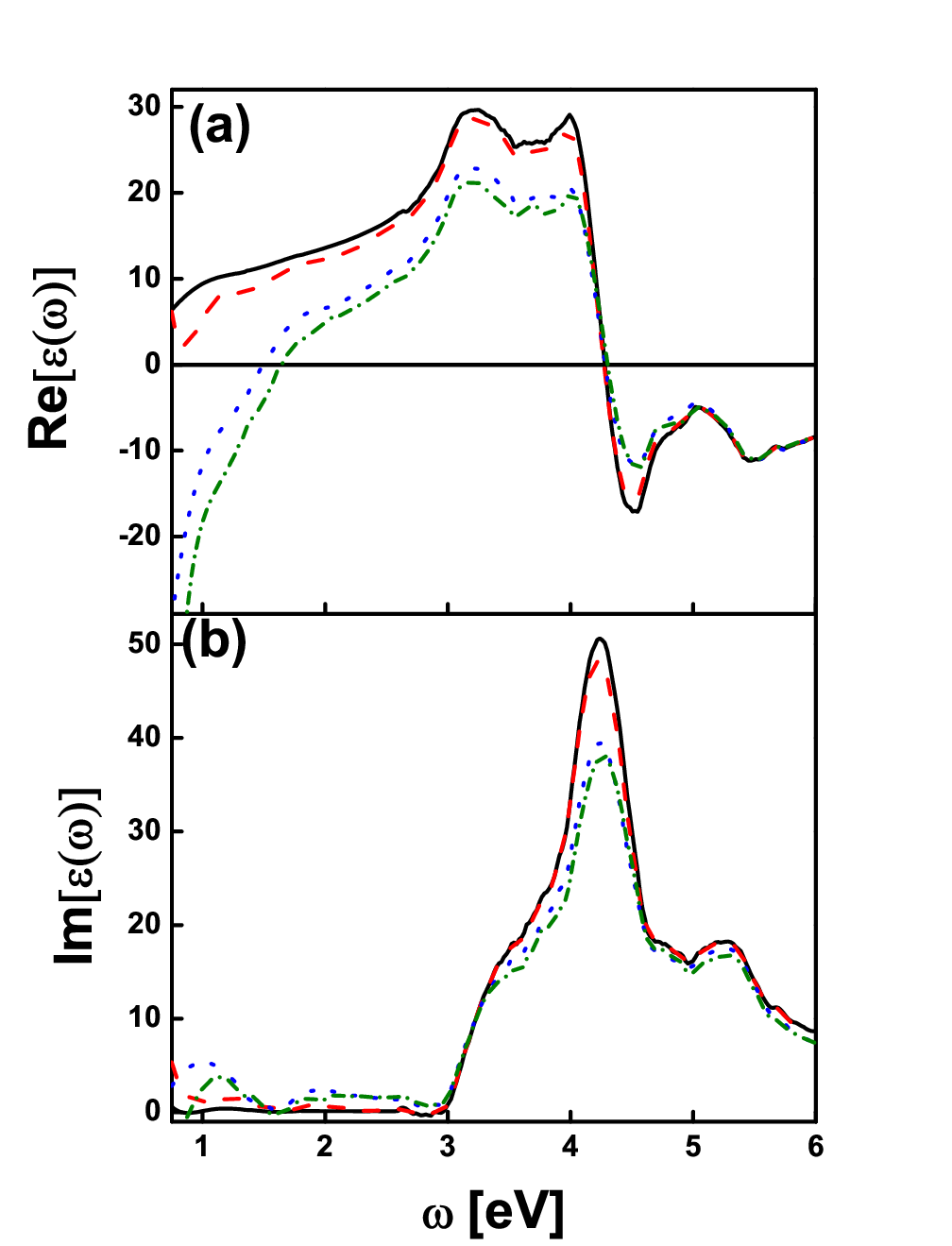}
\caption{Real and imaginary part of the dielectric function of photoexcited silicon irradiated by 90fs laser pulse with wavelength 800 nm, linearly polarized along the [111] crystal direction. In Fig. (a,b) the different line types designate different laser fluences: 1.8 J/cm$^2$ (solid line), 3.2 J/cm$^2$ (dashed line), 3.6 J/cm$^2$ (dotted line) and 4.5 J/cm$^2$ (dashed-dot line).}
\label{fig:Fig7}
\end{figure}

\begin{figure}[h]
\centering\includegraphics[width=0.5\linewidth]{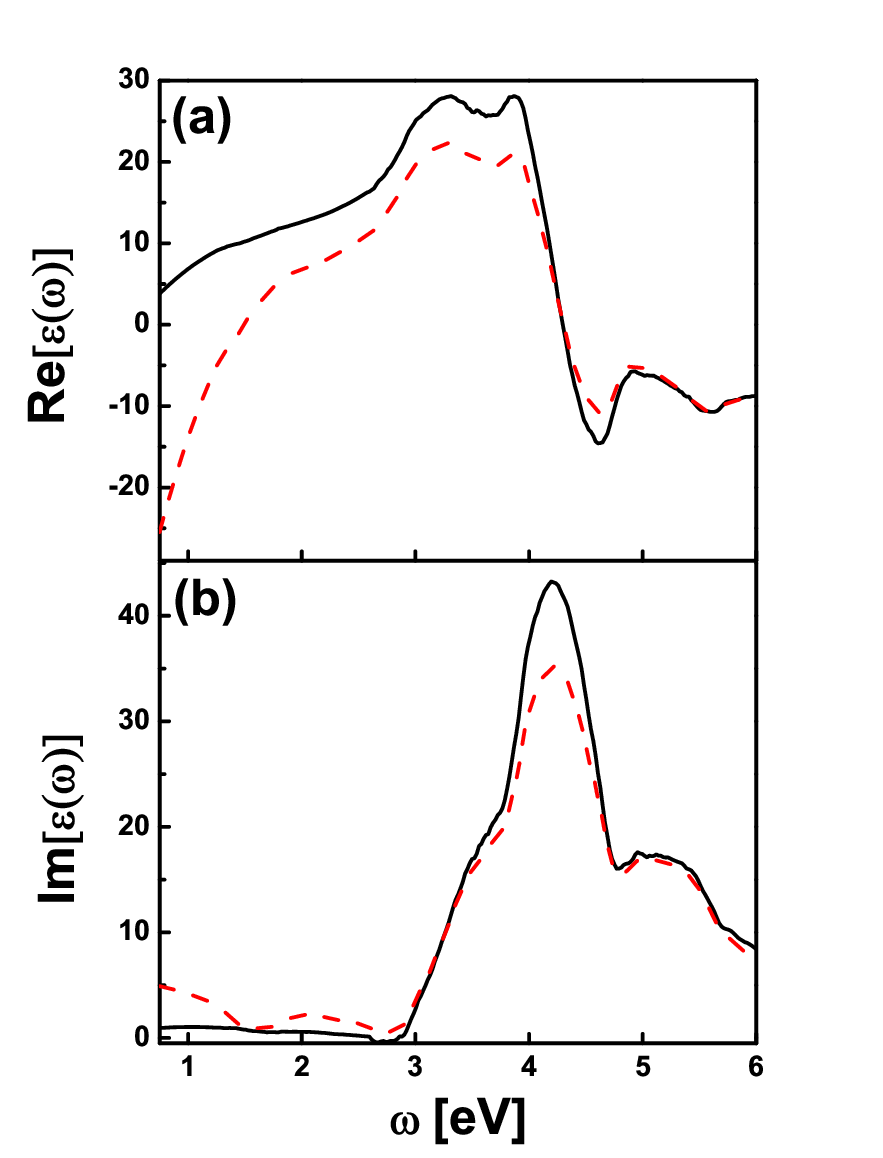}
\caption{Real (a) and  imaginary part (b) of the dielectric function of photoexcited silicon irradiated by 90fs laser pulse with wavelength 800 nm and fluence 3.6 J/cm$^2$. The laser is linearly polarized along the [001] crystal direction (solid line) and along the [111] direction (dotted line).}
\label{fig:Fig7a}
\end{figure}

For high level of electronic excitation, electron-hole pairs modify the dielectric properties of silicon. The real and imaginary parts of the dielectric function of the photoexcited Si are shown in Fig.\ref{fig:Fig7}(a-b). The main feature in the response of the photoexcited solid is the negative divergence of the real part of the dielectric function in the infrared region (cf. Fig.\ref{fig:Fig7}(a)), as a result of plasma formation of free charge carriers. Depending on the laser intensity the real part of the dielectric function vanishes ${{\rm Re}}[\varepsilon(\omega_p)]=0$, and determines the resonance frequency $\omega_p$ of collective bulk plasmon oscillations of the electron gas. As the peak laser intensity increases, the position of the plasmon resonance undergoes a progressive blueshift. Fig.\ref{fig:Fig7}(b) shows also that the imaginary part of the dielectric function becomes appreciable below the direct absorption edge. The diminution of the main absorption peak in the UV region, is accompanied by transfer of spectral weight below the absorption edge and into the Drude peak near $\omega=0$. Similar results were found in Ref.\cite{Sato2014p1}. An interesting feature of the response is that the dielectric tensor has off-diagonal components in the laser-excited state, even though the crystal exhibits cubic symmetry, cf. Fig.\ref{fig:Fig7a} (a-b).

\begin{figure}[h]
\centering\includegraphics[width=1\linewidth]{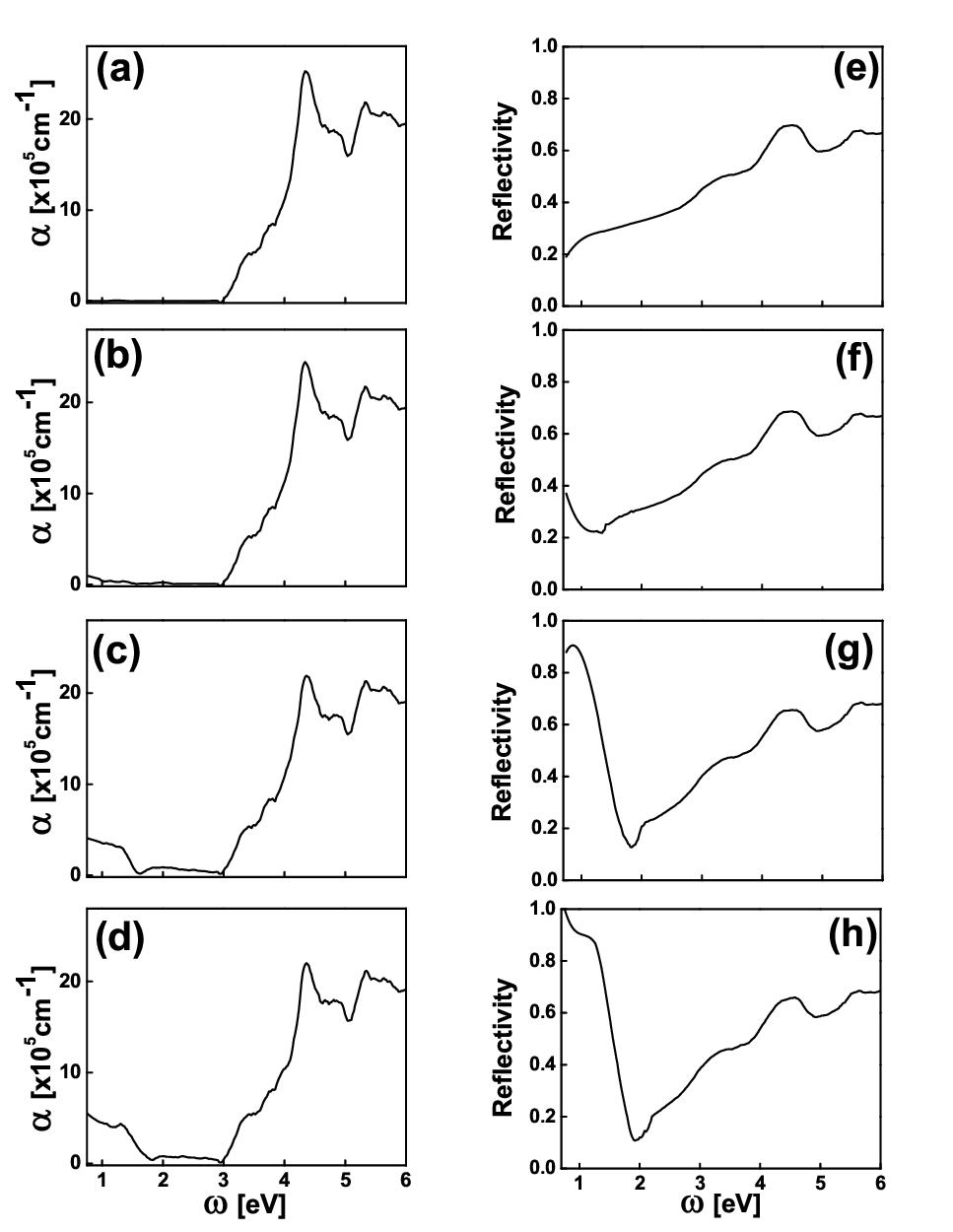}
\caption{Frequency dependent absorption coefficient (left panel) and normal incidence reflectivity (right panel) of photoexcited silicon after irradiation by 90fs laser pulse with wavelength 800 nm, linearly polarized along the [111] crystal direction.
The fluence of the applied laser is 1.8 J/cm$^2$ in Fig.(a), 2.7 J/cm$^2$ in Fig.(b), 3.1 J/cm$^2$ in Fig.(c) and 3.6 J/cm$^2$ in Fig.(d).}
\label{fig:Fig8}
\end{figure}

The frequency-dependent absorption coefficient $\alpha(\omega)$ and the normal-incidence reflectivity $R(\omega)$ of the photo-excited solid are also shown
In Fig.\ref{fig:Fig8}. Here the absorption coefficient is related to the extinction coefficient $\kappa$ of the photo-exicted Si by $\alpha=4 \pi \kappa/ \lambda$, where $\lambda=2 \pi c/\omega$ is the laser wavelength and $c$ is the speed of light in vacuum. For below threshold fluences, the dielectric response of Si is exhibited. The main absorption peak gradually decreases with increasing the peak laser intensity. At the same time, a strongly absorbing state of silicon for near infrared laser wavelengths is created in the breakdown plasma, when the pumped solid becomes opaque. For such large absorption coefficient in the NIR region with $\alpha \sim 10^5 $ cm$^{-1}$ , the laser light probes only a thin surface layer less than 1 $\mu$m in thickness. As a result, the reflectivity of the excited Si becomes very sensitive to the surface quality of the crystal (contamination and oxidation of the surface layer).

\begin{figure}[h]
\centering\includegraphics[width=1\linewidth]{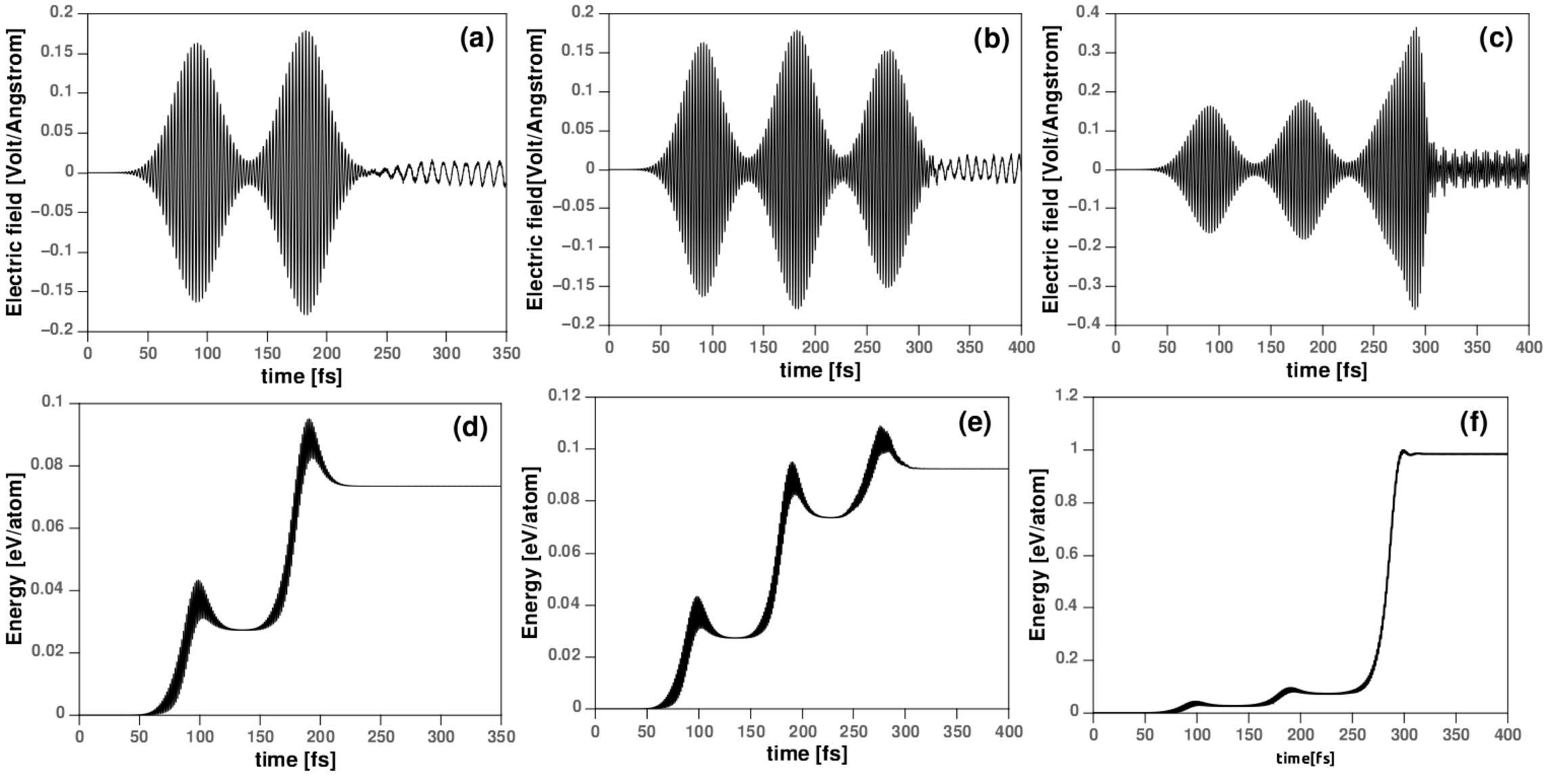}
\caption{Time evolution of the electric field (in V/\AA) of intense femtosecond pulse train transmitted in bulk silicon. In Fig.(a) the fluence of the two pulses is 1.5 J/cm$^2$, in Fig.(b) the fluence of the first two pulses is 1.5 J/cm$^2$ and the fluence of the third pulse is 0.9 J/cm$^2$. In Fig.(c) the three pulses have equal fluence 1.5 J/cm$^2$.  Fig.(d-f) display the corresponding time evolution of the electronic excitation energy in eV per Si atom. The laser pulses in vacuum are linearly polarized along the [001] crystal direction, their wavelength is 800 nm and duration of each one is 30fs.}
\label{fig:tps}
\end{figure}

The effect of the modified dielectric properties can be probed by irradiation with multiple femtosecond laser pulses.
We apply double and triple 30fs pulse sequences with below-threshold fluences,  three different irradiation regimes were considered: i) double pulse sequence: the fluence per pulse is 1.5 J/cm$^2$, ii) triple pulse sequence: the fluences of the first two pulses are equal 1.5 J/cm$^2$, the fluence of the third pulse is $0.9$ J/cm$^2$ and iii) the crystal is subjected to three identical pulses with fluence per pulse $\Phi=$ 1.5 J/cm$^2$. The results are presented in Fig.\ref{fig:tps}(a-f). For the double pulse sequence i) the first transmitted pulse is weakly distorted by the laser-matter interaction and dielectric response of Si is exhibited: the  pulse exhibits temporary Gaussian profile with electric field  $\ec=\ec_{{\rm ext}} /\epsilon_0$, where $\epsilon_0=12$  is the static bulk dielectric constant of silicon. The peak field strength inside the bulk reaches 0.15 V/\AA, which ionizes valence electrons by multi-photon transitions and deposits 0.03 eV per Si atom. The dielectric constant at the laser wavelength is slightly decreased by free-carrier polarization created by the first pulse. Therefore the electric field of the second transmitted pulse inside the bulk increases and reaches 0.18 V/\AA~at its peak. As a result, the probability for photionization increases and this second pulse deposits 0.05 eV per Si atom. The cumulative effect of the two pulses manifests in creation of weak self-sustained electric field after conclusion of the second pulse. Thus electron-hole pairs born in valence and conduction bands are organized in self-sustained oscillation with plasma frequency $\omega_p=0.45$ eV, which is well below the one-photon energy.

For the triple sequence in case ii), the first two pulses has
ionized the medium and produced weak plasma oscillations,  but
the plasma frequency does not match the laser oscillation frequency of the weak third pulse. Because the dielectric constant of Si has decreased after the conclusion of the previous two pulses, the peak field strength of the third pulse inside the bulk reaches 0.15 V/\AA and produces more electronic excitation, consequently the plasma oscillation frequency increases to $\omega_p=0.6$ eV. The net energy deposited by the pulse train is nearly 0.1 eV/atom and is slightly below the thermal melting threshold of Si.
In case iii), the fluence of the third pulse is increased,
its temporal profile is strongly distorted by the non-linear polarization of the solid: the rising edge of the pulse
increases linearly with time, the electric field strength increases rapidly by more than 0.1 V/\AA~during each 10fs, until it reaches 0.4 V\AA~ at the pulse peak. The pulse produces
strong electronic excitation and practically diminishes the dielectric constant at the laser wavelength, i.e. the laser frequency matches the plasma oscillation frequency, whence resonant transfer of energy to electrons and optical breakdown occur. After the breakdown, the conduction electrons display metallic response and screen out the laser electric field: a steep edge is formed in the back of the third pulse and a frequency up-chirp is exhibited.

The incident laser pulses not only ionize the medium and thus modify the dielectric properties of the photoexcited solid, but also create a coherent superposition of states in the valence and conduction bands, which continously exchange energy with the laser field. For near threshold fluences, wave-functions of valence electrons are strongly distorted and include admixtures from anti-bonding orbitals in the conduction band, whence resonant one-photon transitions into vacant higher-energy conduction band states occur with high probability. Thus  when optical breakdown in Si occurs, the covalent bonds in Si are softened by the strong electric fields (with strength $\approx$ 2 V/\AA~inside vacuum and 0.2 V/\AA~inside the bulk), these softened bonds ionize very efficiently via one-photon absorption.

\begin{figure}[h]
\centering\includegraphics[width=1\linewidth]{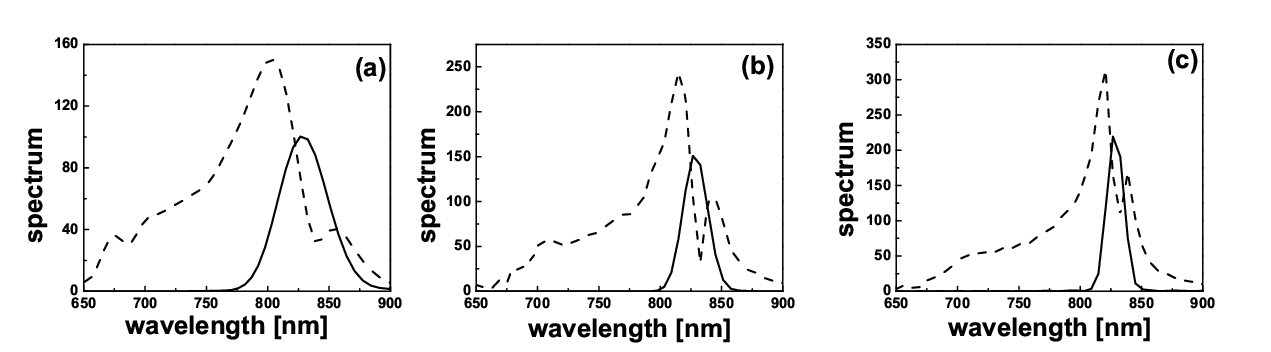}
\caption{Frequency spectrum of a transmitted pulse in bulk silicon irradiated by pulsed laser with near-infrared wavelength 800 nm. Different line types in Figs.(a-b) designate different laser energy fluences. In Fig.(a) the pulse duration is 30 fs, the fluence is 1.5 J/cm$^2$ (solid line) and  3 J/cm$^2$ (dashed line). In Fig. (b) the pulse duration is 60 fs and the fluence is 1.8 J/cm$^2$ (solid line) and 5.4 J/cm$^2$ (dashed line) and in Fig.(c) the pulse length is 90 fs, the fluence is 2.7 J/cm$^2$ (solid line) and 8.1 J/cm$^2$ (dashed line). In Fig.(a-c) the applied laser is linearly polarized along the [001] crystal direction.}
\label{fig:Fig5c}
\end{figure}

The spectral profile of a single pulse with below and above threshold fluence is shown in Fig.\ref{fig:Fig5c}. For below threshold fluences, the spectrum is centered around the laser wavelength and exhibits Gaussian profile. For above threshold fluences, a broad pedestal appears on the blue side of the central wavelength and extends into the visible region. This is consistent with plasma model of superbroadening proposed in Ref. \cite{Bloembergen1973,Yablonovitch1974}: the generation of free carriers creates dense breakdown plasma in the wake of the pulse, whence the refractive index of Si decreases (and becomes less than 1) and contributes to self-phase modulation in the transmitted pulse. As a result the laser oscillation frequency becomes time-dependent and upshifts in the back of the pulse, which causes severe spectral broadening and manifests in broad supercontinuum generation.

\subsection{Bonding properties of photo-excited Si}

\begin{figure}[h]
\centering\includegraphics[width=1\linewidth]{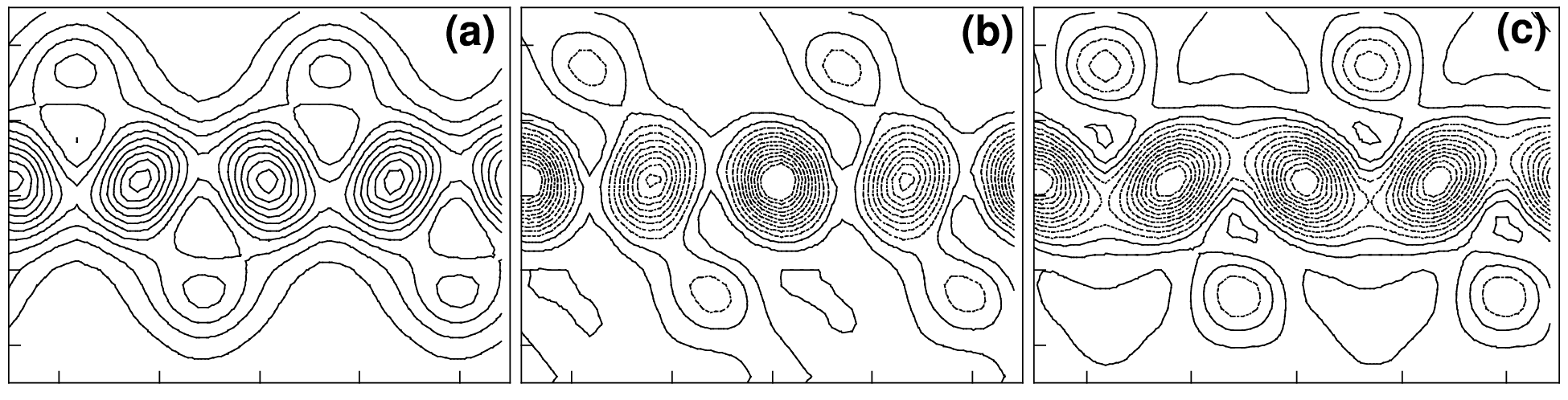}
\caption{Contour plots of the electron number density in the
(110) plane in silicon. Fig.(a) - ground state electron density. Fig.(b) and Fig.(c) display the local density difference map $n(\rs)-n_0(\rs)$ after irradiation of Si by 30fs laser pulse with wavelength 800 nm and fluence 3 J/cm$^2$. In Fig.(b) and Fig.(c) the laser is linearly polarized along the [111] and [001] crystal directions, respectively.  In Figs. (b,c), negative contours (charge depletion) are indicated by the dotted lines and positive contours (charge accumulation) by the solid lines. The contour interval is 0.01 a.u. in Fig.(a), and is 0.0005 a.u. in Fig.(b) and Fig.(c).
 }
\label{fig:Fig9}
\end{figure}

Fig. \ref{fig:Fig9}(b-c) display the change of electron density distribution in the (110) plane in Si after photo-excitation by intense 30fs laser pulse. For comparison, Fig.\ref{fig:Fig9}(a) shows the ground state valence charge density map characterized by bond charges located in the middle of each Si-Si bond. The local pseudo-potential model produces closed elliptic density contour lines with long axis perpendicular to the bonding direction. Photo-ionization causes redistribution of charge density, electrons which occupy bonding sites move into the interstitial anti-binding regions. This leads to weakening of the Si-Si bonds, the amplitude of the bond charge  decreases and thus the bond order is reduced, cf. Fig. \ref{fig:Fig9}(b-c). More electrons are displaced from bonds parallel to the laser polarization direction ([111] direction) as compared to bonds that are oriented perpendicularly ([1 $\bar{1}$ 1] direction). The bond charges remain as centers of inversion of the diamond lattice, their amplitude decreases and shape changes.

There is simple criteria for stability of the diamond lattice, cf.
\cite{Heine1976,Kudryashov2002}. The transverse acoustic (TA) phonons in Si are unstable when the bond charge vanishes \cite{Martin1969}, the crystal no longer resists to shear stress and melts. Because the bond charge in Si is $Z_b=2/\epsilon_0=1/6$ \cite{Phillips1968},  the creation of a single electron-hole pair destroys the bond charge $Z_b$ in six Si-Si bonds, thus  the excitation of electron-hole pairs of number density $n_c$ softens the transverse acoustic phonon modes by
\begin{equation}
\omega_{{\rm TA}}^2 \rightarrow \omega_{{\rm TA}}^2 \left(1 - \epsilon_0 \frac{n_c}{n_0} \right)
\end{equation}
where $n_0=32/a_0^3$ is the average bulk electron density. The simple estimation predicts that when 8 \% of the valence electrons are promoted into the conduction band, the diamond lattice becomes unstable and melts on ultrashort time scale. This is consistent with both our result and more detailed calculation based on phenomenological bond-charge model in Ref. \cite{Biswas1982}.

Charge transfer from bonding sites results in filling of vacuum and interstitial regions, which partially destroys the covalent bonding interaction between the Si atoms. The freed electrons create an uncompensated outward kinetic pressure, such that crystal is no longer in mechanical equilibrium and should expand uniformly.  However this lattice expansion is slow on a ultrashort time scale, because the speed of sound in Si $c_s=10^4$ m/s. Ref.\cite{Stampfli1990} modeled the elasticity of the photo-excited Si using a phenomenological valence force field model with three adjustable parameters - central force constant, bond-bending constant and a non-equilibrium photo-induced pressure constant. They considered shear distortions of the diamond lattice and showed that large internal pressure makes negative the elastic shear modulus, softens the transverse acoustic phonons and results in ultra-fast melting of silicon.

\begin{figure}[h]
\centering\includegraphics[width=1\linewidth]{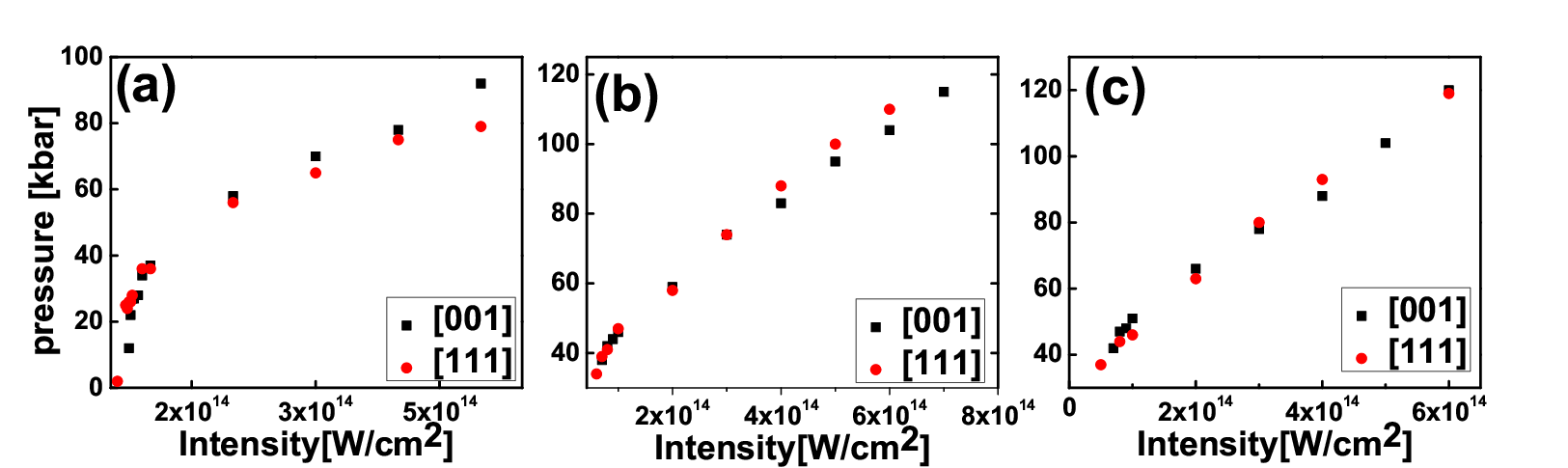}
\caption{Laser intensity dependence of the photoinduced pressure (in kbar) in crystalline Si after the irradiation by pulsed laser of near-infrared wavelength 800 nm. The  pulse duration is 30 fs in Fig.(a), 60 fs in Fig.(b) and 90 fs in Fig.(c). Different symbols designate different laser polarization directions - [001] axis (squares) and the [111] axis (circles).  }
\label{fig:Fig10}
\end{figure}

Our independent calculation of the change of the macroscopic stress tensor of Si for fixed-ion positions \cite{Nielsen1985}, shows that large uncompensated  positive pressure arises after photo-excitation and that more than 80 \% of the photoinduced pressure is due increase of the electronic kinetic energy. The electronic kinetic pressure is evaluated from
\begin{equation}
p = \frac{1}{3} \sum_{v,\gc} \int_{{\rm BZ}} \frac{d^3 \ks}{4 \pi^3}  \left(
|c_{v,\gc+\ks}|^2 - |c^0_{v,\gc+\ks}|^2 \right) (\gc+\ks)^2
\end{equation}
where $\gc$ are the reciprocal-lattice wave-vectors, the quasi-momentum $\ks$ extends over the first Brillouin zone,
$v$ labels initially occupied valence band states,
$c^0_{v,\gc+\ks}$ and $c_{v,\gc+\ks}$ are the electronic momentum-space wave-functions before and after the excitation. Fig.\ref{fig:Fig10} shows the dependence of the photo-induced pressure on the laser intensity. For high level of electronic excitation $n_c \ge 10^{22}$ cm$^{-3}$, the pressure $p \sim$ 10 GPa exceeds the elastic strength of Si. More importantly, $p$ exhibits a characteristic sub-linear scaling trend with $p \sim I^{5/9}$
and because $n_c \sim I^{1/3}$, the famous relation between  pressure and density $p \sim n_c^{5/3}$ for degenerate Fermi gas is obtained. Thus the Pauli's exclusion principle is prominent in the breakdown regime. This result differs from the prediction of the phenomenological model in Ref.\cite{Stampfli1990}, which obtains pressure increasing linearly with the increase of conduction electron density. The breakdown plasma gives rise to  large Fermi pressure, that can not be compensated by electron-ion attraction. Thus electron emission from the silicon surface occurs, leaving behind positive holes. The charged Si surface may become electrostatically unstable, and emit massive positively charged projectiles on a longer time scale,.

\section{Conclusion}

The optical breakdown threshold of crystalline silicon is obtained as a function of the laser energy fluence, pulse duration and polarization direction. The occurrence of breakdown is associated
with plasma formation and highly efficient free-carrier absorption. The transmitted pulse is distorted and becomes subject to self-steepening and self-phase modulation, the corresponding breakdown field strength is self-limited and does not exceed 0.4 V/\AA. The severe spectral broadening of transmitted pulse results in generation of supercontinuum extending into the visible spectral region. The optical properties of Si undergo irreversible modification in the infrared spectral region: both absorption and reflectivity of near infrared wavelengths are greatly enhanced in the breakdown plasma. In this regime of high electronic excitation, an unbalanced photo-induced degeneracy pressure results in mechanical instability of the lattice, which may be considered as a precursor to ultrafast melting and ablation of Si. We plan
more detailed investigation of the elasticity of the photoexcited solid for a follow-up paper.

\section*{Acknowledgements}
Support from the Bulgarian National Science Fund under Contract No. KP-06-KOST is acknowledged. This research is based on work supported by the Air Force Office of Scientific Research under award number FA9550-19-1-7003.

\clearpage

\clearpage

\end{document}